\documentclass[twocolumn,showpacs,preprintnumbers]{revtex4}
\usepackage{amssymb}
\usepackage{amsmath}
\usepackage{graphicx}
\usepackage{dcolumn}
\usepackage{bm}

\begin{document}

\title{Bipartite Concurrence and Localized Coherence}
\author{Chang-shui Yu}
\email{quaninformation@sina.com; ycs@dlut.edu.cn}
\author{He-shan Song}
\email{hssong@dlut.edu.cn}
\affiliation{School of Physics and Optoelectronic Technology, Dalian University of
Technology,\\
Dalian 116024, P. R. China}
\date{\today }

\begin{abstract}
Based on a proposed coherence measure, we show that the local
coherence of a bipartite quantum pure state (coherence of its
reduced density matrix) is exactly the same as the minimal average
coherence with all potential pure-state realizations under
consideration. In particular, it is shown that bipartite concurrence
of pure states just captures the maximal difference between local
coherence and the average coherence of one subsystem induced by
local operations on the other subsystem with the assistance of
classical communications, which provides an alternative operational
meaning for bipartite concurrence of pure states. The relation
between concurrence and the proposed coherence measure can also be
extended to bipartite mixed states.
\end{abstract}

 \maketitle

\section{\textbf{INTRODUCTION}}

Coherence and entanglement arise from quantum superposition, the most
distinctive and puzzling feature of quantum mechanics. Quantum coherence is
an important subject in quantum mechanics, where decoherence due to the
interaction with an environment is a crucial issue that is of fundamental
interest. If there exists coherence among multiple quantum subsystems, a
special nonlocal coherence---quantum entanglement, may be generated besides
the local coherence of each constituent subsystem. As an ingredient of
quantum mechanics, quantum entanglement has been recognized to be an
important physical resource in quantum information processing including
quantum communication and quantum computation and plays a key role in
quantum information theory [1-4]. Recently, many works based on some special
models have been made to show the relation between local decoherence and
disentanglement of a composite quantum system by considering the interaction
with environments [5-10]. In fact, so long as there exists interactions
between two subsystems, the coherence of each subsystem might also be
changed. For example, if a composite quantum system is maximally entangled,
each subsystem is completely incoherent. It is natural to ask how the
entanglement is related to the local coherence.

In fact, the above question means finding some kind of operational meaning
of the entanglement measure that we will employ. Even though quantification
of entanglement has attracted many interests in recent years and a lot of
entanglement measures have been proposed and explored from different
viewpoints [11-17], an entanglement measure is usually concerned mainly
about the monotonicity under local operation and classical communication
(LOCC), i.e., not increase under LOCC operations [18-21], hence only a few
entanglement measures have been considered from the operational meanings
point of view [22-27]. The most popular two examples are entanglement cost
[22,23] and distillable entanglement [23-25] which show the conversion rate
between the entangled state of interests and the maximally entangled state.
As a remarkable entanglement measure, concurrence [16] has been widely
employed in lots of cases of quantum information theory. However, to our
knowledge, concurrence \textit{per se} of pure states is related to the
purity of one subsystem which only roughly or qualitatively shows the role
of the other subsystem [28].

In this paper, we focus on the relation between concurrence and
localized coherence, which can provide an alternative operational
meaning for concurrence. Suppose Alice and Bob share a composite
bipartite state, Alice's local coherence is determined by her
reduced density matrix but independent of its pure-state
realization. However, if Bob performs some operations on his
subsystem, with the assistance of classical communication Alice
might owe her quantum ensemble with different average local
coherence. For example, for a Bell state in $\sigma _{z\text{
}}$representation,
Alice's reduced density matrix is completely mixed. But if Bob performs a $%
\sigma _{x\text{ }}$measurement on his subsystem and tell Alice his outcome,
Alice will obtain a pure state with maximal coherence. In this sense, we say
that the coherence can be localized assisted by Bob. In this paper we
propose a coherence measure by collecting contribution of all off-diagonal
elements of a density matrix. Based on this coherence measure, we show that
the local coherence of a bipartite pure state is just the same as the
minimal average coherence with all potential pure-state realizations taken
into consideration. In particular, it is shown that with this coherence
measure, concurrence can be regarded as the difference between the maximal
and the minimal localized (local) coherence. Thus it provides an operational
meaning for concurrence. This is much like what we have found for $\left(
2\otimes 2\otimes n\right) $-dimensional 3-tangle which can be considered as
the difference between the concurrence of assistance and concurrence of $%
\left( 2\otimes 2\right) $-dimensional subsystem [29]. This paper is
organized as follows. In Sec. II, we consider the coherence measure of
quantum systems of a qubit and show the relation between the coherence
measure and the concurrence of $(2\otimes n)$-dimensional quantum systems;
In Sec. III, we focus on coherence measure of high-dimensional quantum
system and consider the relation between the coherence measure and the
concurrence of a general $(n_{1}\otimes n_{2})$-dimensional quantum systems;
In Sec. IV, we extend both the relations given in Sec II and III to
concurrence of bipartite mixed states. The conclusion is drawn in Sec. V.

\section{Quantum coherence of qubit and concurrence of $(2\otimes n)$
dimensional pure states}

\subsection{Quantification of coherence}

It has been shown that a good definition of coherence does not only depend
on the state of the system $\rho $, but also depend on the alternatives
under consideration which are usually attached to different eigenvalues of
an observable $A$. Since the off-diagonal elements of $\rho $ characterize
interference, they are usually called \textit{coherences} with respect to
the basis in which $\rho $ is written [30-32]. The measurements on the
observables that do not commute with $A$ can reveal the interference. It is
obvious that if $\rho $ is diagonalized, there is not any relevant
coherences with respect to that basis. Thus one can straightforwardly
quantify the coherence in given basis by measuring the distance between the
quantum state $\rho $ and the nearest incoherent state.

\textbf{Definition 1:} If $\rho $ is written in some basis, the coherence
with respect to the same basis can be measured by
\begin{equation}
D(\rho )=\left\vert \left\vert \rho -\sigma ^{\ast }\right\vert \right\vert
_{1}=\sum\limits_{i\neq j}\left\vert \rho _{ij}\right\vert ,
\end{equation}%
where $\sigma ^{\ast }$ is the diagonal matrix with $\sigma _{ii}^{\ast
}=\rho _{ii}$ and $\left\vert \left\vert \cdot \right\vert \right\vert _{1}$
is the "Entrywise" norm. In fact, $\left\vert \left\vert \cdot \right\vert
\right\vert _{1}$ can also be replaced by Frobenius norm $\left\vert
\left\vert \cdot \right\vert \right\vert _{F}$ for some convenient
applications.

It is easily to find that that $D(\rho )=\min_{\sigma \in \mathcal{I}%
}\left\vert \left\vert \rho -\sigma \right\vert \right\vert _{1}=\left\vert
\left\vert \rho -\sigma ^{\ast }\right\vert \right\vert _{1}$ where $%
\mathcal{I}$ is the set of incoherent states with the same basis to $\rho $.
This shows the direct geometric meaning of the coherence measure. In
addition, the measure collects the contribution of all off-diagonal elements
of $\rho $ which is consistent with what we have stated previously.

\subsection{Localizable coherence}

There exists infinitely many pure-state realization of a given mixed state.
Unlike quantum entanglement of a bipartite quantum state $\rho $ which is
defined as the minimal average entanglement with all pure-state realizations
of $\rho $ taken into account, in usual it seems not to be meaningful to
define the average coherence of a mixed state by considering the different
pure-state realizations. However, it is not the case if we have known that $%
\rho _{A}$ owned by Alice was reduced from a bipartite state $\varrho _{AB}$
shared with Bob, i.e., $\rho _{A}=Tr_{B}\varrho _{AB}$. Based on GHJW
theorem [33], any pure-state realization of $\rho _{A}$ can be obtained by
appropriate POVM performed on subsystem $B$ [34]. In this sense, if Bob
informs Alice of the measurement outcomes via classical communication, Alice
can obtain the corresponding pure state $\left\vert \phi _{i}\right\rangle $
with probability $p_{i}$. In other words, Alice will obtain the
corresponding coherence $D(\left\vert \phi _{i}\right\rangle )$ with
probability $p_{i}$. Averagely, the coherence that Alice can obtain should
be given by%
\begin{equation}
\bar{D}(\rho _{A})=\sum\limits_{i}p_{i}D(\left\vert \phi _{i}\right\rangle ).
\end{equation}%
In this case, $D(\rho _{A})$ defined in eq. (1) is called local
coherence because it describes the coherence of the local subsystem
$A$ in contrast to the whole composite system $\varrho _{AB}$, and
the average coherence given in eq. (2) can also be called localized
coherence because the average coherence is generated based on Bob's
assistance.

\textbf{Definition 2:} The localizable coherence of $\rho _{A}$ is
defined as the maximal average coherence with all possible
pure-state realizations taken into account$,$i.e.,
\begin{equation}
D_{L}(\rho _{A})=\max \bar{D}(\rho _{A}).
\end{equation}%
It is implied in the definition that one can distinguish the different
pure-state realizations with the help of LOCC between the two components $A$
and $B$ of the composite quantum system $\varrho _{AB}$.

\subsection{Relation between coherence and concurrence}

\textbf{Theorem 1.} Suppose $\mathcal{E}=\{p_{i},\left\vert \psi
_{i}\right\rangle \}$ is a potential pure-state realization of a
quantum state of qubit $\rho $, then the coherence measure
\begin{eqnarray}
D(\rho ) &=&\sum\limits_{i\neq j}\left\vert \rho _{ij}\right\vert =\min_{%
\mathcal{E}}\bar{D}(\rho _{A})  \notag \\
&=&\min_{\mathcal{E}}\sum_{i}p_{i}D\left( \left\vert \psi _{i}\right\rangle
\right) =\lambda _{1}-\lambda _{2}
\end{eqnarray}%
and the localizable coherence
\begin{equation}
D_{L}(\rho )=\max_{\mathcal{E}}\sum_{i}p_{i}D\left( \left\vert \psi
_{i}\right\rangle \right) =\lambda _{1}+\lambda _{2},
\end{equation}%
where $\lambda _{i}$ is the square root of the eigenvalues of $\rho \sigma
_{x}\rho ^{\ast }\sigma _{x}$ and $\sigma _{x}=\left(
\begin{array}{cc}
0 & 1 \\
1 & 0%
\end{array}%
\right) $.

\textbf{Proof.} At first, by a simple algebra, one can easily find that eq.
(1) can be rewritten as $D(\left\vert \psi \right\rangle )=\left\vert
\left\langle \psi ^{\ast }\right\vert \sigma _{x}\left\vert \psi
\right\rangle \right\vert $ for a pure state of qubit. Thus for a mixed
state $\rho =\sum p_{i}\left\vert \psi _{i}\right\rangle \left\langle \psi
_{i}\right\vert $, the average coherence can be given by%
\begin{equation}
\bar{D}\left( \rho \right) =\sum_{i}p_{i}\left\vert \left\langle \psi
_{i}^{\ast }\right\vert \sigma _{x}\left\vert \psi _{i}\right\rangle
\right\vert .
\end{equation}%
Considering the matrix notation $\rho =\Psi W\Psi ^{\dag }$, where the
columns of $\Psi $ correspond to $\left\vert \psi _{i}\right\rangle $ and $W$
is a diagonal matrix with diagonal entries corresponding to $p_{i}$, one can
find that

\begin{equation}
\bar{D}\left( \rho \right) =\sum_{i}\left\vert W^{1/2}\Psi ^{T}\sigma
_{x}\Psi W^{1/2}\right\vert _{ii}
\end{equation}%
with superscript $T$ denoting transpose operation. Based on the eigenvalue
decomposition: $\rho =\Phi M\Phi ^{\dag }$, where the columns of $\Phi $
correspond to the eigenvectors and $M$ is a diagonal matrix with diagonal
entries corresponding to the eigenvalues, it is easily find that $%
W^{1/2}\Psi =U^{T}\Phi ^{T}M^{1/2}$ with $UU^{\dag }=\mathbf{1}$ and $%
\mathbf{1}$ the identity. Thus eq. (7) can be rewritten as%
\begin{equation}
\bar{D}\left( \rho \right) =\sum_{i}\left\vert U^{T}M^{1/2}\Phi ^{T}\sigma
_{x}\Phi M^{1/2}U\right\vert _{ii}.
\end{equation}%
The minimal and maximal (localizable coherence) average coherence can be
directly calculated from eq. (8) based on Thompson Theorem [35,36] and Ref.
[37]. Namely,%
\begin{eqnarray}
\bar{D}^{\min }\left( \rho \right)  &=&\min_{U}\sum_{i}\left\vert
U^{T}M^{1/2}\Phi ^{T}\sigma _{x}\Phi M^{1/2}U\right\vert _{ii}  \notag \\
&=&\lambda _{1}-\lambda _{2},
\end{eqnarray}%
and%
\begin{eqnarray}
D_{L}\left( \rho \right)  &=&\max_{U}\sum_{i}\left\vert U^{T}M^{1/2}\Phi
^{T}\sigma _{x}\Phi M^{1/2}U\right\vert _{ii}  \notag \\
&=&\lambda _{1}+\lambda _{2},
\end{eqnarray}%
where $\lambda _{i}$ is the singular values of matrix $M^{1/2}\Phi
^{T}\sigma _{x}\Phi M^{1/2}$ in decreasing order or the square roots of the
eigenvalues of $\rho \sigma _{x}\rho ^{\ast }\sigma _{x}$.

In order to explicitly show the $\bar{D}^{\min }\left( \rho \right) $ and $%
D_{L}\left( \rho \right) $, we suppose $\rho =\left(
\begin{array}{cc}
a & b^{\ast } \\
b & c%
\end{array}%
\right) $, where $a$ and $c=1-a$ are real and $ac-\left\vert b\right\vert
^{2}\geqslant 0$ due to the positive $\rho $. From eq. (1), it is obvious
that the coherence of $\rho $ is $D(\rho )=2\left\vert b\right\vert .$
Substitute $\rho $ into eq. (9), we can obtain that
\begin{equation}
\rho \sigma _{x}\rho ^{\ast }\sigma _{x}=\left(
\begin{array}{cc}
ac+\left\vert b\right\vert ^{2} & 2ab^{\ast } \\
2bc & ac+\left\vert b\right\vert ^{2}%
\end{array}%
\right) .
\end{equation}%
The eigenvalue equation of $\rho \sigma _{x}\rho ^{\ast }\sigma _{x}$ can be
given by
\begin{equation}
\Lambda ^{2}-2(ac+\left\vert b\right\vert ^{2})\Lambda +\left( ac-\left\vert
b\right\vert ^{2}\right) ^{2}=0.
\end{equation}%
Thus based on Vieta's Theorem [38], one can easily find that
\begin{equation}
\bar{D}^{\min }\left( \rho \right) =\lambda _{1}-\lambda _{2}=2\left\vert
b\right\vert =D(\rho ),
\end{equation}%
and
\begin{equation}
D_{L}\left( \rho \right) =\lambda _{1}+\lambda _{2}=2\sqrt{ac}.
\end{equation}%
In particular, eq. (13) shows that $\bar{D}^{\min }\left( \rho
\right) $ is exactly the same as the coherence of $\rho $. In this
sense, we can redescribe the coherence of $\rho $ as the minimal
average coherence.$\hfill\Box$

\textbf{Theorem 2. }For a bipartite $(2\otimes n)-$dimensional quantum pure
state $\left\vert \varphi \right\rangle _{AB}$ with $\rho
_{A}=Tr_{B}\left\vert \varphi \right\rangle _{AB}\left\langle \varphi
\right\vert $ defined in 2 dimension, the concurrence $C\left( \left\vert
\varphi \right\rangle _{AB}\right) $ of $\left\vert \varphi \right\rangle
_{AB}$ satisfies
\begin{equation}
C^{2}\left( \left\vert \varphi \right\rangle _{AB}\right) =\bigskip
D_{L}^{2}\left( \rho _{A}\right) -D^{2}\left( \rho _{A}\right) .
\end{equation}
 \textbf{Proof.} Suppose the reduced density matrix of the bipartite
pure state $\left\vert \varphi \right\rangle _{AB}$ is given by
\begin{equation}
\rho _{A}=Tr_{B}\left\vert \varphi \right\rangle _{AB}\left\langle \varphi
\right\vert =\left(
\begin{array}{cc}
a & b^{\ast } \\
b & c%
\end{array}%
\right) ,
\end{equation}%
then the concurrence of $\left\vert \varphi \right\rangle _{AB}$ is defined
[39] as%
\begin{equation}
C(\left\vert \varphi \right\rangle _{AB})=\sqrt{2\left( 1-Tr\rho
_{A}^{2}\right) }.
\end{equation}%
Substitute eq. (16) into eq. (17), one can have
\begin{equation}
C(\left\vert \varphi \right\rangle _{AB})=\sqrt{4\left( ac-\left\vert
b\right\vert ^{2}\right) }.
\end{equation}%
Based on eq. (13) and eq. (14), it is obvious that
\begin{equation}
\bigskip D_{L}^{2}\left( \rho _{A}\right) -D^{2}\left( \rho _{A}\right)
=4\left( ac-\left\vert b\right\vert ^{2}\right) .
\end{equation}%
Therefore, eq. (15) holds.$\hfill\Box$

\section{Quantum coherence of qudit and concurrence of general bipartite
pure states}

\subsection{Quantification of coherence and localizable coherence for a qudit%
}

For a high-dimensional quantum state, the key question is how to generalize
the coherence measure and the average coherence of quantum qubit states. The
discussion in Section II provides a direct understanding of average
coherence, especially for qubit systems. In a different matter, we can give
a new understanding to average coherence of high dimensional quantum system.
Since coherence is closely related to the nonzero off-diagonal elements, it
requires at least two levels for a given quantum system (for example, the
excited and ground states of an atom) in order to demonstrate the coherence.
In other words, a two-level system can be considered as the minimal unit in
researching coherence, which just corresponds to two off-diagonal elements
of density matrix in terms of \textbf{Definition 1}. In this sense, if $\rho
_{AB}$ is shared by Alice and Bob, Alice can be only concerned about the
coherence with respect to the given basis in some $2\times 2$ subspace and
then collect all the contributions of different subspace.

For an $n$-dimensional density matrix $\rho _{A}$, there exist $N=\frac{%
n(n-1)}{2}$ alternative $2\times 2$ subspace. The quantum state in each $%
2\times 2$ subspace can be achieved by
\begin{equation}
\rho _{i}=\frac{L_{i}\rho _{A}L_{i}^{\dagger }}{TrL_{i}\rho
_{A}L_{i}^{\dagger }},
\end{equation}%
where $Tr\left( L_{i}\rho _{A}L_{i}^{\dagger }\right) $ is normalization
factor and $L_{i}$ is the $2\times n$ matrix which can be derived from $%
\left\vert S\right\vert $ ($\left\vert \cdot \right\vert $ denotes the
absolute value of the matrix elements) by deleting the row where all the
elements are zero with $S$ denoting the generator of the group $SO(n)$. The
average coherence in the $i$th subspace can be given by $\bar{D}(\rho _{i})$
defined as eq. (2). Define an $N-$ dimensional average coherence vector as
\begin{equation}
\mathcal{D}(\rho _{A})=[\bar{D}(\rho _{1}),\bar{D}(\rho _{2}),\cdot \cdot
\cdot ,\bar{D}(\rho _{N})]
\end{equation}%
and the corresponding weight-like vector as
\begin{equation}
\mathcal{P}(\rho _{A})=[TrL_{1}\rho _{A}L_{1}^{\dagger },TrL_{2}\rho
_{A}L_{2}^{\dagger },\cdot \cdot \cdot ,TrL_{N}\rho _{A}L_{N}^{\dagger }],
\end{equation}%
then the total average coherence of all subspace can be defined as the
length of the weighted vector, i.e.,%
\begin{equation}
\bar{D}_{F}(\rho _{A})=\left\Vert \mathcal{P\circ D}\right\Vert ,
\end{equation}%
where $\mathcal{\circ }$ denotes the Hadamard product and $\left\Vert \cdot
\right\Vert $ denotes the $L_{2}$ norm of a vector and the subscript $F$
will be explained later. It is obvious that $\bar{D}_{F}(\rho _{A})$ and $%
\mathcal{D}(\rho _{A})$ depend on Bob's operations. In this sense, we can
define a vector of maximal average coherence as
\begin{eqnarray}
\mathcal{D}_{L}(\rho _{A}) &=&\left\Vert \mathcal{P\circ }\max \mathcal{D}%
\right\Vert   \notag \\
&=&[\bar{D}_{L}(\rho _{1}),\bar{D}_{L}(\rho _{2}),\cdot \cdot \cdot ,\bar{D}%
_{L}(\rho _{N})]
\end{eqnarray}%
with $\bar{D}_{L}(\cdot )$ is given by eq. (3), by which we can analogously
define the localizable coherence as follows.

\textbf{Definition 3:} The localizable coherence of $\rho _{A}$ is
defined as
the length of the weighted maximal average coherence vector $\mathcal{D}%
_{L}(\rho _{A}),$i.e.,
\begin{equation}
D_{FL}(\rho _{A})=\left\Vert \mathcal{P\circ D}_{L}\right\Vert .
\end{equation}

At the end of this subsection, we would like to emphasize that the
generalized coherence measures $\bar{D}_{F}(\rho _{A})$ and $D_{FL}(\rho
_{A})$ can be reduced to $\bar{D}(\rho _{A})$ and $D_{L}(\rho _{A})$,
respectively, when $\rho _{A}$ is a density matrix of a qubit. We have shown
that $D(\rho _{A})=\min_{\mathcal{E}}\bar{D}(\rho _{A})$ for a qubit density
matrix $\rho _{A}$, the analogous relation with $\bar{D}_{F}(\rho _{A})$ and
$D_{F}(\rho _{A})$ taken into account is also satisfied for a
high-dimensional $\rho _{A}$, which will be proved in the next subsection.
In addition, it should be noted that the subscripts $F$ means that $%
D_{F}(\rho _{A})=\sqrt{\sum\limits_{i\neq j}\left\vert \rho
_{Aij}\right\vert ^{2}}$, namely, in \textbf{Definition 1 }of coherence
measure, we employ Frobenius norm.

\subsection{Relation between coherence and concurrence}

\textbf{Theorem 3.} For a quantum state of qudit $\sigma $, let $\mathcal{D}%
(\sigma )$ be the average coherence vector defined as eq. (21) and $\mathcal{%
D}_{L}(\sigma )$ be the maximal average coherence with the corresponding
weight-like vector $\mathcal{P}(\sigma )$ defined as eq. (22). Then the the
coherence measure $D_{F}(\sigma )$ can be given by
\begin{eqnarray}
D_{F}(\sigma ) &=&\sqrt{\sum\limits_{i\neq j}\left\vert \sigma
_{ij}\right\vert ^{2}}=\left\Vert \mathcal{P}(\sigma )\mathcal{\circ }\min
\mathcal{D}(\sigma )\right\Vert  \notag \\
&=&\sqrt{\sum_{j}\left( \tilde{\lambda}_{1}^{j}-\tilde{\lambda}%
_{2}^{j}\right) ^{2}}
\end{eqnarray}%
and the localizable coherence
\begin{eqnarray}
D_{FL}(\sigma ) &=&\left\Vert \mathcal{P}(\sigma )\mathcal{\circ D}%
_{L}(\sigma )\right\Vert  \notag \\
&=&\sqrt{\sum_{j}\left( \tilde{\lambda}_{1}^{j}+\tilde{\lambda}%
_{2}^{j}\right) ^{2}},
\end{eqnarray}%
where $\tilde{\lambda}_{k}^{j}$ is the square root of the eigenvalues of $%
\rho \left\vert S_{j}\right\vert \rho ^{\ast }\left\vert S_{j}\right\vert $.

\textbf{Proof.} Let $\mathcal{E}=\{q_{i},\left\vert \chi _{i}\right\rangle \}
$ be a potential decomposition of $n$-dimensional density matrix $\sigma $.
Substitute $\mathcal{E}$ into eq. (26) (or eq. (23)), one can find that
\begin{gather}
TrL_{j}\sigma L_{j}^{\dagger }\bar{D}(\frac{L_{j}\sigma L_{j}^{\dagger }}{%
TrL_{j}\sigma L_{j}^{\dagger }})  \notag \\
=\sum_{i}q_{i}D\left( L_{j}\left\vert \chi _{i}\right\rangle \left\langle
\chi _{i}\right\vert L_{j}^{\dagger }\right) =\sum_{i}q_{i}\left\vert
\left\langle \chi _{i}^{\ast }\right\vert L_{j}^{T}\sigma
_{x}L_{j}\left\vert \chi _{i}\right\rangle \right\vert   \notag \\
=\sum_{i}\left\vert \tilde{U}^{T}\tilde{M}^{1/2}\tilde{\Phi}^{T}\left\vert
S_{j}\right\vert \tilde{\Phi}\tilde{M}^{1/2}\tilde{U}\right\vert _{ii},
\end{gather}%
where $\tilde{U}\tilde{U}^{\dag }=\mathbf{1}$ by which any decomposition of $%
\sigma =\tilde{\Psi}\tilde{W}\tilde{\Psi}^{\dag }$ is related to the
eigenvalue decomposition $\sigma =\tilde{\Phi}\tilde{M}\tilde{\Phi}^{\dag }$%
. Based on Thompson theorem and Ref. [37], one can find that
\begin{eqnarray}
TrL_{j}\sigma L_{j}^{\dagger }\min_{\mathcal{E}}(\frac{L_{j}\sigma
L_{j}^{\dagger }}{TrL_{j}\sigma L_{j}^{\dagger }}) &=&\tilde{\lambda}%
_{1}^{j}-\tilde{\lambda}_{2}^{j}, \\
TrL_{j}\sigma L_{j}^{\dagger }\max_{\mathcal{E}}(\frac{L_{j}\sigma
L_{j}^{\dagger }}{TrL_{j}\sigma L_{j}^{\dagger }}) &=&\tilde{\lambda}%
_{1}^{j}+\tilde{\lambda}_{2}^{j},
\end{eqnarray}%
where $\tilde{\lambda}_{k}^{j}$ is the square root of the eigenvalues of $%
\sigma \left\vert S_{j}\right\vert \sigma ^{\ast }\left\vert
S_{j}\right\vert $ in decreasing order. In eq. (29) and eq. (30), it should
be emphasized that $\sigma \left\vert S_{j}\right\vert \sigma ^{\ast
}\left\vert S_{j}\right\vert $ has only two nonzero eigenvalues ($\tilde{%
\lambda}_{1}^{j}\ $and $\tilde{\lambda}_{2}^{j}$), since the nonzero block
of $\sigma \left\vert S_{j}\right\vert \sigma ^{\ast }\left\vert
S_{j}\right\vert $ is completely the same as $L_{j}\sigma L_{j}^{\dagger
}\sigma _{x}L_{j}\sigma ^{\ast }L_{j}^{T}\sigma _{x}$. Thus
\begin{eqnarray}
\left\Vert \mathcal{P\circ }\min \mathcal{D}\right\Vert  &=&\sqrt{%
\sum_{j}\left( \tilde{\lambda}_{1}^{j}-\tilde{\lambda}_{2}^{j}\right) ^{2}},
\\
\left\Vert \mathcal{P\circ }\max \mathcal{D}\right\Vert  &=&\left\Vert
\mathcal{P}(\sigma )\mathcal{\circ D}_{L}(\sigma )\right\Vert =\sqrt{%
\sum_{j}\left( \tilde{\lambda}_{1}^{j}+\tilde{\lambda}_{2}^{j}\right) ^{2}}.
\end{eqnarray}

In fact, one can find that for each $L_{j}$, $L_{j}\sigma L_{j}^{\dagger }$
can be written by
\begin{equation}
L_{j}\sigma L_{j}^{\dagger }=\left(
\begin{array}{cc}
\sigma _{kk} & \sigma _{kl} \\
\sigma _{kl}^{\ast } & \sigma _{ll}%
\end{array}%
\right) ,
\end{equation}%
where $\sigma _{kk}$, $\sigma _{ll}$ are the $k$th and $l$th diagonal
elements of $\sigma $ and $\sigma _{kl}$ is the off-diagonal element of $%
\sigma $ subject to the two diagonal elements. Analogous to the proof of
\textbf{Theorem 1}, one can find that
\begin{equation}
\begin{array}{c}
\tilde{\lambda}_{1}^{j}-\tilde{\lambda}_{2}^{j}=2\left\vert \sigma
_{kl}\right\vert , \\
\tilde{\lambda}_{1}^{j}+\tilde{\lambda}_{2}^{j}=2\sqrt{\sigma _{kk}\sigma
_{ll}}.%
\end{array}%
\end{equation}%
Since each pair of off-diagonal elements of $\sigma $ corresponds to a $L_{j}
$, the contribution of all the off-diagonal elements can be described as%
\begin{eqnarray}
D_{F}(\sigma ) &=&\sqrt{\sum\limits_{i\neq j}\left\vert \sigma
_{ij}\right\vert ^{2}}  \notag \\
&=&\sqrt{\sum_{j}\left( \tilde{\lambda}_{1}^{j}-\tilde{\lambda}%
_{2}^{j}\right) ^{2}}=\left\Vert \mathcal{P\circ }\min \mathcal{D}%
\right\Vert .
\end{eqnarray}%
Eqs. (31,35) and eq. (32) show that this theorem holds.$\hfill\Box$

\textbf{Theorem 4.} For a bipartite $(n_{1}\otimes n_{2})-$dimensional
quantum pure state $\left\vert \eta \right\rangle _{AB}$ with $\sigma
_{A}=Tr_{B}\left\vert \eta \right\rangle _{AB}\left\langle \eta \right\vert $
defined in $n_{1}$ dimension, the concurrence $C\left( \left\vert \eta
\right\rangle _{AB}\right) $ of $\left\vert \eta \right\rangle _{AB}$
satisfies
\begin{equation}
C^{2}\left( \left\vert \eta \right\rangle _{AB}\right) =\bigskip
D_{FL}^{2}\left( \sigma _{A}\right) -D_{F}^{2}\left( \sigma _{A}\right) .
\end{equation}

\textbf{Proof. }Since the concurrence of $\left\vert \eta \right\rangle
_{AB} $ is defined as eq. (17), based on $\sigma _{ij}$---the entries of $%
\sigma _{A} $, $C\left( \left\vert \eta \right\rangle _{AB}\right) $ can be
rewritten by%
\begin{equation}
C\left( \left\vert \eta \right\rangle _{AB}\right) =\sqrt{%
4\sum\limits_{ij}\left( \sigma _{ii}\sigma _{jj}-\left\vert \sigma
_{ij}\right\vert ^{2}\right) }.
\end{equation}%
According to eq. (31) and eq. (32), we have%
\begin{eqnarray}
&&D_{FL}^{2}\left( \sigma _{A}\right) -D_{F}^{2}\left( \sigma _{A}\right)
\notag \\
&=&\sum_{j}\left( \tilde{\lambda}_{1}^{j}+\tilde{\lambda}_{2}^{j}\right)
^{2}-\sum_{j}\left( \tilde{\lambda}_{1}^{j}-\tilde{\lambda}_{2}^{j}\right)
^{2}.
\end{eqnarray}%
Substitute eq. (34) into eq. (36), one can find that
\begin{eqnarray}
&&D_{FL}^{2}\left( \sigma _{A}\right) -D_{F}^{2}\left( \sigma _{A}\right)
\notag \\
&=&4\sum_{kl}\left( \sigma _{kk}\sigma _{ll}-\left\vert \sigma
_{kl}\right\vert ^{2}\right) .
\end{eqnarray}%
Comparing eq. (37) with eq. (39), one can conclude that eq. (36)
holds.$\hfill\Box$

\section{Quantum coherence and bipartite concurrence of mixed states}

In this section, we will show that \textbf{Theorem 2} and \textbf{Theorem 4}
can be extended to bipartite mixed states. For a bipartite mixed state $%
\sigma _{AB}$, one can always introduce an auxiliary system $C$ such that $%
\left\vert \psi \right\rangle _{ABC}$ is a purification of $\sigma _{AB}$.
If subsystem $A$ is 2-dimensional, based on \textbf{Theorem 2} one can
obtain
\begin{equation}
C^{2}\left( \left\vert \psi \right\rangle _{A\left( BC\right) }\right)
=\bigskip D_{L}^{2}\left( \sigma _{A}\right) -D^{2}\left( \sigma _{A}\right)
.
\end{equation}%
If subsystem $A$ is more than 2-dimensional, based on \textbf{Theorem 4} one
can obtain
\begin{equation}
C^{2}\left( \left\vert \psi \right\rangle _{A\left( BC\right) }\right)
=\bigskip D_{FL}^{2}\left( \sigma _{A}\right) -D_{F}^{2}\left( \sigma
_{A}\right) .
\end{equation}%
In eq. (40) and eq. (41), $\sigma _{A}=Tr_{BC}\left\vert \psi \right\rangle
_{ABC}\left\langle \psi \right\vert $. Since concurrence $C\left( \left\vert
\psi \right\rangle _{A\left( BC\right) }\right) $ is an entanglement
monotone-----it does not increase under LOCC [18,19] and $\sigma _{AB}$ can
always be obtained from $\left\vert \psi \right\rangle _{A\left( BC\right) }$
by local operations on subsystem $C$, one has
\begin{equation}
C\left( \left\vert \psi \right\rangle _{A\left( BC\right) }\right) \geqslant
C\left( \sigma _{AB}\right) .
\end{equation}%
Thus we can have the following theorem.

\textbf{Theorem 5: } For bipartite mixed state $\sigma _{AB}$, if subsystem $%
A$ is 2 dimensional, the concurrence satisfies%
\begin{equation}
\bigskip C^{2}\left( \sigma _{AB}\right) \leq D_{L}^{2}\left( \sigma
_{A}\right) -D^{2}\left( \sigma _{A}\right) ,
\end{equation}%
otherwise,
\begin{equation}
\bigskip C^{2}\left( \sigma _{AB}\right) \leq D_{FL}^{2}\left( \sigma
_{A}\right) -D_{F}^{2}\left( \sigma _{A}\right) .
\end{equation}

\section{Conclusion and discussion}

In summary, we have shown that the local coherence based on a proposed
coherence measure can be understood as the minimal average coherence with
all potential pure-state realizations taken into account. In particular, we
have revealed the relation between the local coherence including localizable
coherence and bipartite concurrence of pure states which  provides an
alternative operational meaning for concurrence of pure states. In addition,
it is also shown that the relation can also be extended to the case of
bipartite mixed state.

Before the end, we would like to briefly discuss the potential applications
of our relations. As mentioned in Introduction, a lot of works have been
done to study disentanglement and local decoherence by considering different
$(2\otimes 2)-$dimensional physical models. However, for high-dimensional
quantum systems, there does not generally exist an analytic entanglement
measure which greatly limits the relevant researches. It can be easily found
that the coherence measures presented in this paper can be analytically
calculated, in particular the relations given in \textbf{Theorem 5} provide
an upper bound of concurrence, therefore, one can find that a sufficient
condition of disentanglement can be provided by local decoherence.

{\textbf{V. ACKNOWLEDGEMENT}}

This work was supported by the National Natural Science Foundation of China,
under Grant No. 10805007, No. 10875020, and the Doctoral Startup Foundation
of Liaoning Province.


\begin{thebibliography}{99}
\bibitem{[28]} Concurrence of a bipartite pure state $\left\vert \phi
\right\rangle _{AB}$ can be written by $C(\left\vert \phi \right\rangle
_{AB})=\sqrt{2(1-Tr\rho _{A}^{2})}$ [39] with $\rho _{A}=Tr_{B}\left\vert
\phi \right\rangle _{AB}\left\langle \phi \right\vert $, where $\sqrt{%
2(1-Tr\rho _{A}^{2})}$(or $1-Tr\rho _{A}^{2}$ equivalently, ) is usually
used to quantify the purity of $\rho _{A}$. For some decomposition $\rho
_{A}=\sum\limits_{i}p_{i}\sigma _{A}^{i}$, $C(\left\vert \phi \right\rangle
_{AB})=\sqrt{2(1-Tr\left[ \sum\limits_{i}p_{i}\sigma _{A}^{i}\right] ^{2})}%
\geqslant \sum\limits_{i}p_{i}\sqrt{2(1-Tr\left[ \sigma _{A}^{i}\right] ^{2})%
}$ based on the convacity, where the "=" holds if and only $\sigma _{A}^{i}$
is the same for any $i,$ but this condition is trivial. In particular, when $%
\sigma _{A}^{i}$ is pure for all $i$,\ the above inequality only provides a
trivial lower bound $0$. In this sense, one can not find the explicit role
of subsystem B.
\end{thebibliography}
\end{document}